\documentclass[12pt]{article}
\usepackage{epsf}

\begin{document}

\newcommand{\gtrsim}{ \mathop{}_{\textstyle \sim}^{\textstyle >} }
\newcommand{\lesssim}{ \mathop{}_{\textstyle \sim}^{\textstyle <} }

\newcommand{\rem}[1]{{\bf #1}}

\renewcommand{\thefootnote}{\fnsymbol{footnote}}
\setcounter{footnote}{0}
\begin{titlepage}

\def\thefootnote{\fnsymbol{footnote}}

\begin{center}

\hfill TU-599\\
\hfill RCNS-00-01\\
\hfill hep-ph/0007328\\
\hfill July, 2000\\

\vskip .75in

{\Large \bf 
CP violation in $B_d\rightarrow \phi K_S$ in SUSY GUT\\ 
with right-handed neutrinos
}

\vskip .75in

{\large
  Takeo Moroi
}

\vskip 0.25in

{\em
Department of Physics, Tohoku University,
Sendai 980-8578, JAPAN}

\end{center}

\vskip .5in

\begin{abstract}

    CP violation in the $B_d$ system is discussed in the
    supersymmetric grand unified theory (GUT) with the right-handed
    neutrinos.  Above the GUT scale, the right-handed down-type
    squarks couple to the right-handed neutrinos.  Due to the
    renormalization group effect, flavor violations in the lepton
    sector may be transfered to the right-handed down-type squark mass
    matrix, which affects the CP violation in the $B$ decay.  Taking
    into account this effect, we compare the CP violation in
    $B_d\rightarrow \psi K_S$ and $B_d\rightarrow \phi K_S$ processes.
    We will find that a significant difference is possible between the
    CP violating phases in two decay processes.

\end{abstract}
\end{titlepage}

\renewcommand{\thepage}{\arabic{page}}
\setcounter{page}{1}
\renewcommand{\thefootnote}{\#\arabic{footnote}}
\setcounter{footnote}{0}

Many efforts have been made to understand the CP violation.  In
particular, experimentally, $B$-factories are now being to measure the
CP violation in the $B$ system and to obtain insights into the physics
behind the CP violation.  One of the most important purposes of the
$B$-factories is to check the unitarity of the
Cabibbo-Kobayashi-Maskawa (CKM) matrix $V_{\rm CKM}$ using the
$B_d\rightarrow\psi K_S$ decay mode; in the standard model, the phase
${\rm arg}(-[V_{\rm CKM}]_{cd}[V_{\rm CKM}^*]_{cb}/ [V_{\rm
CKM}]_{td}[V_{\rm CKM}^*]_{tb})$ is determined with
$B_d\rightarrow\psi K_S$ \cite{hph9908520}.  Since the phase in the
CKM matrix is the only source of the CP violation in the standard
model,\footnote{We neglect the strong CP problem.} such a measurement
will provide severe constraints on the CP violation in other
processes.

There can be, however, extra contribution to the CP violation from a
new physics beyond the standard model.  Since the standard model
suffers from the so-called hierarchy problem, i.e., the stability of
the Higgs mass parameter against the radiative corrections, we are
forced to introduce a new physics at the electroweak scale.
Furthermore, the deficits of the atmospheric and solar neutrino fluxes
strongly suggest non-vanishing masses and mixings in the neutrino
sector.  Since the new physics is, in general, CP violating, the
$B$-factories may be able to observe phenomena related to the new CP
violating interactions.

In this letter, we consider supersymmetric grand unified theory (GUT)
with the right-handed neutrinos as the new physics.  In this model,
the hierarchy problem is solved by the cancellation of the quadratic
divergences between bosonic and fermionic loops due to the
supersymmetry (SUSY), while the neutrino masses are generated by the
seesaw mechanism\ \cite{seesaw}.  In such a model, the right-handed
down-type squarks interact with the right-handed neutrinos and colored
Higgs.  Of course, such interactions are negligible for the low energy
physics at the tree level since the colored Higgs is as heavy as the
GUT scale.

However, they become important through the renormalization group (RG)
effect \cite{rgeffect}.  (Similar effects have been studied for the
low energy flavor violating processes
\cite{fv,JHEP0003-019,hph0002142}.)  In Ref.\ \cite{JHEP0003-019}, it
was discussed that the right-handed neutrinos affects the structure of
the flavor and CP violations in the mass matrix of the right-handed
down-type squarks in SUSY GUT.\footnote{In Ref.\ \cite{hph0002142},
similar effect was also discussed.  However, in this article, one of
the most important effects, i.e., the effect of the phases in the GUT
models, was neglected.} In particular, SUSY contribution to the
$\epsilon_K$ parameter can be as large as the currently measured value
of $\epsilon_K$ if we adopt an $O(1)$ phase in the neutrino Yukawa
matrix.  In this letter, we consider different effects, that is, CP
violation in the decay processes of the $B$-mesons.  We will see that
the SUSY contribution to the decay amplitude of the $B_d\rightarrow
\phi K_S$ process can be large, and that the phase in the decay
amplitude can be as large as $O(0.1)$ which is much larger than the
standard model prediction. We will emphasize that phases in the GUT
Lagrangian play a crucial role for this enhancement.

Let us first introduce the model we consider.  To make our point
clearer, we consider a SUSY $SU(5)$ model with the right-handed
neutrinos.  The superpotential of the model is
\begin{eqnarray}
    W_{\rm GUT} = \frac{1}{8} \Psi_i \left[Y_{U}\right]_{ij} \Psi_j H
    + \Psi_i \left[Y_{D}\right]_{ij} \Phi_j \bar{H} 
    + N_i \left[Y_{N}\right]_{ij} \Phi_j H
    + \frac{1}{2} N_i \left[M_{N}\right]_{ij} N_j,
\label{W_GUT}
\end{eqnarray}
where $\Psi_i$, $\Phi_i$, and $N_i$ are ${\bf 10}$, ${\bf \bar{5}}$,
and singlet matter fields of $SU(5)$ in $i$-th generation, while
$\bar{H}$ and $H$ are Higgs fields which are in ${\bf \bar{5}}$ and
${\bf 5}$ representations, respectively.\footnote{We assume that other
chiral multiplets, in particular, ones which are responsible for the
$SU(5)$ breaking, do not affect the following argument.} Here, $M_N$
is the Majorana mass matrix for the right-handed neutrinos, and for
simplicity, we adopt the universal structure:
\begin{eqnarray}
    \left[M_{N}\right]_{ij} = M_{\nu_R} \delta_{ij}.
\label{M_nR}
\end{eqnarray}
The Yukawa matrix $Y_{U}$ is a complex symmetric matrix while $Y_{D}$
and $Y_{N}$ are complex matrix.  We take the basis where the Yukawa
matrices become
\begin{eqnarray}
    Y_U = V_Q^T \hat{\Theta}_Q \hat{Y}_U V_Q,~~~
    Y_D = \hat{Y}_D,~~~
    Y_N = \hat{Y}_N V_L \hat{\Theta}_L,
\end{eqnarray}
where $\hat{Y}$'s are real diagonal matrices:
\begin{eqnarray}
    \hat{Y}_U = {\rm diag} (y_{u},  y_{c},  y_{t}),~~~
    \hat{Y}_D = {\rm diag} (y_{d},  y_{s},  y_{b}),~~~
    \hat{Y}_N = {\rm diag} (y_{\nu_1},  y_{\nu_2},  y_{\nu_3}),
\end{eqnarray}
while $\hat{\Theta}$'s are diagonal phase matrices:
\begin{eqnarray}
    \hat{\Theta}_Q = 
    {\rm diag}
    (e^{i\phi^{(Q)}_1}, e^{i\phi^{(Q)}_2}, e^{i\phi^{(Q)}_3}),~~~
    \hat{\Theta}_L = 
    {\rm diag}
    (e^{i\phi^{(L)}_1}, e^{i\phi^{(L)}_2}, e^{i\phi^{(L)}_3}),
\end{eqnarray}
where phases obey the constraints
$\phi^{(Q)}_1+\phi^{(Q)}_2+\phi^{(Q)}_3=0$ and
$\phi^{(L)}_1+\phi^{(L)}_2+\phi^{(L)}_3=0$.  Furthermore, $V_Q$ and
$V_L$ are unitary mixing matrices parameterized by three mixing angles
and one CP violating phase.

The quarks and leptons in the standard model are embedded in the
$SU(5)$ multiplets as
\begin{eqnarray}
    \Psi_i \simeq
    \{ 
    Q,\ V_Q^\dagger \hat{\Theta}_Q^\dagger \bar{U},\ 
    \hat{\Theta}_L \bar{E} 
    \}_i,~~~
    \Phi_i \simeq \{ \bar{D},\ \hat{\Theta}_L^\dagger L \}_i,
\label{embed}
\end{eqnarray}
where $Q_i ({\bf 3},{\bf 2})_{1/6}$, $\bar{U}_i ({\bf \bar{3}},{\bf
1})_{-2/3}$, $\bar{D}_i ({\bf \bar{3}},{\bf 1})_{1/3}$, $L_i ({\bf
1},{\bf 2})_{1/2}$, and $\bar{E}_i ({\bf 1},{\bf 1})_{1}$ are quarks
and leptons in $i$-th generation with the $SU(3)_C\times SU(2)_L\times
U(1)_Y$ gauge quantum numbers as shown.  With the embedding
(\ref{embed}), the superpotential for the light particles is
\begin{eqnarray}
    W_{\rm SSM} &=&
    Q_i \left[ V_Q^T \hat{Y}_U \right]_{ij} \bar{U}_j H_u
    + Q_i \left[ \hat{Y}_D \right]_{ij} \bar{D}_j H_d
    + \bar{E}_i \left[ \hat{Y}_E \right]_{ij} L_j H_d
    + N_i \left[ \hat{Y}_N V_L \right]_{ij} L_j H_u
    \nonumber \\ &&
    + \frac{1}{2} M_{\nu_R} N_i N_i,
\label{W_SSM}
\end{eqnarray}
with $H_u$ and $H_d$ being the up- and down-type Higgs fields,
respectively.  In the $SU(5)$ limit, $\hat{Y}_E=\hat{Y}_D$ although
this relation does not hold for the first and second generations.  We
expect that some mechanism, like higher dimensional operator
suppressed by the Planck scale, fixes this problem.  We assume such a
mechanism does not affect the following analysis.  In Eq.\ 
(\ref{W_SSM}), the unitary matrix $V_Q$ becomes the CKM matrix:
$V_Q\simeq V_{\rm CKM}$.  Furthermore, the neutrino mass matrix after
the seesaw mechanism is given by
\begin{eqnarray}
[m_{\nu_L}]_{ij} = \frac{\langle H_u\rangle^2}{M_{\nu_R}} 
\left[ V_L^T \hat{Y}_N^2 V_L \right]_{ij}
= \frac{v^2\sin^2\beta}{2M_{\nu_R}} 
\left[ V_L^T \hat{Y}_N^2 V_L \right]_{ij},
\label{nuL_mass}
\end{eqnarray}
where $v\simeq 246~{\rm GeV}$, and $\beta$ parameterizes the relative
size of the Higgs vacuum expectation values: $\tan\beta\equiv\langle
H_u\rangle /\langle H_d\rangle$.  Therefore, with Eq.\ (\ref{M_nR}),
$V_L$ plays the role of the neutrino mixing matrix \cite{PTP28-870}.
In this letter, we mainly consider the neutrino mass matrix suggested
by the atmospheric neutrino flux deficit \cite{n2k_Sobel}, and also by
the large angle MSW solution to the solar neutrino problem
\cite{n2k_Suzuki}:
\begin{eqnarray}
m_{\nu} \simeq
( \begin{array}{ccc}
0, & 0.004~{\rm eV}, & 0.06~{\rm eV}
\end{array} ),~~~
V_L \simeq
\left( \begin{array}{ccc}
0.91 & -0.30 & 0.30 \\
0.42 & 0.64 & -0.64 \\
0 & 0.71 & 0.70
\end{array} \right).
\label{VL_la}
\end{eqnarray}
Notice that, due to Eq.\ (\ref{nuL_mass}), the neutrino Yukawa
couplings $y_\nu$'s increase as $M_{\nu_R}$ becomes larger with the
fixed light neutrino masses.  With a too large $M_{\nu_R}$, the
neutrino Yukawa coupling becomes non-perturbative below the Planck
scale.  With Eq.\ (\ref{VL_la}), it happens when $M_{\nu_R}\gtrsim
10^{15}\ {\rm GeV}$.  Therefore, we only consider the cases with
$M_{\nu_R}\lesssim 10^{15}\ {\rm GeV}$.

With the embedding (\ref{embed}), all the GUT scale phases drop off
from the Yukawa interactions among the light fields.  However, the
phases $\phi^{(L)}$ and $\phi^{(Q)}$ remain in the colored Higgs
vertices.  Above the GUT scale $M_{\rm GUT}$, interaction among $N$,
$\bar{D}$, and $H_C$ is effective in the following form
\begin{eqnarray}
    W_{\rm GUT} =
    N_i \left[ \hat{Y}_N V_L \hat{\Theta}_L \right]_{ij} \bar{D}_j H_C
    + \cdots
    = \sum_{i,j} y_{\nu_i} \left[V_L\right]_{ij} e^{i\phi^{(L)}_j} 
    N_i \bar{D}_j H_C + \cdots.
\label{W_NDHc}
\end{eqnarray}
Although the effect of the colored Higgs is negligible to the $B$
physics at the tree level, the interaction (\ref{W_NDHc}) is
potentially important since it affects the soft SUSY breaking mass
parameters through the RG effect.  The simplest way to estimate the
effect is to assume the universality of the scalar mass at the reduced
Planck scale $M_*\simeq 2.4\times 10^{18}\ {\rm GeV}$.  This
corresponds to adopting the model so-called mSUGRA.  With this
framework, we will estimate the possible size of the SUSY contribution
to the CP violation in the $B$ system.

In this approach, the soft SUSY breaking parameters are parameterized
by four free parameters: the universal scalar mass $m_0$, the
universal $A$-parameter $a_0$, the gaugino mass $m_{G5}$, and the
$B$-parameter which is determined by the condition for the proper
electroweak symmetry breaking.  With the universal scalar mass at the
reduced Planck scale, we run all the parameters down to the
electroweak scale and evaluate physical quantities using the
parameters at the electroweak-scale.  Of course, the scalar mass
parameters may be non-universal at the reduced Planck scale, and if so
our results can be modified.  Since the off-diagonal elements of the
scalar mass matrix is the most important for our discussion, however,
we expect that the mSUGRA approach gives us a reasonable and
conservative estimation in most of the cases.  The CP violation in the
$B$ system we will discuss will be more enhanced if the non-universal
contribution is larger than the running effect.  If two contributions
are comparable, the signal can be smaller due to an accidental
cancellation.  However, such a cancellation requires a tuning of the
parameters, and we neglect this possibility.

\begin{figure}
\centerline{\epsfxsize=0.5\textwidth\epsfbox{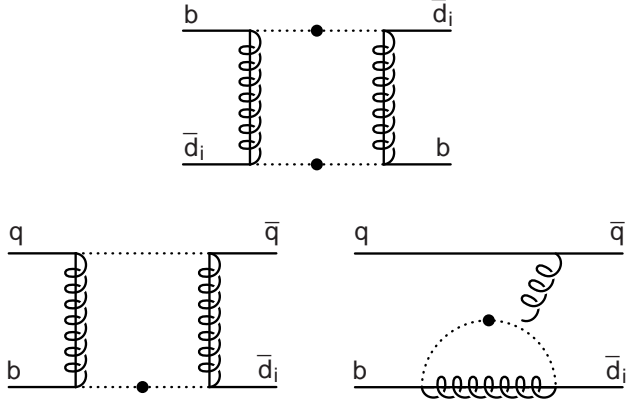}}
\caption{Feynman diagrams contributing to the $\Delta B=2$  and 
$\Delta B=1$ processes. The solid lines are quarks with $d_i=d$ and
$s$, the dashed ones squarks, the wiggled ones with straight lines
gluinos, and the wiggled line is the gluon.  The ``dot'' $\bullet$ on
the squark lines represents the mass insertion. For the box diagrams,
diagrams with crossing gluino lines also exist.}
\label{fig:feyn}
\end{figure}

Next, we estimate the size of the flavor violating off-diagonal
elements in the down-type squark mass matrix.  Relevant part of the
soft SUSY breaking terms is given by
\begin{eqnarray}
    {\cal L}_{\rm soft} &=& 
    - [m^2_{\tilde{Q}}]_{ij} \tilde{Q}_i \tilde{Q}_j^*
    - [m^2_{\tilde{\bar{D}}}]_{ij} 
    \tilde{\bar{D}}_i \tilde{\bar{D}}_j^*
    - \left( [A_D]_{ij} \tilde{Q}_i \tilde{\bar{D}}_j H_d 
        + {\rm h.c.} \right)
    \nonumber \\ &&
    - \frac{1}{2} \left( m_{G3} \tilde{G} \tilde{G}
        + {\rm h.c.} \right),
    \label{L_soft}
\end{eqnarray}
where $\tilde{Q}_i$ and $\tilde{\bar{D}}_i$ are scalar components of
the corresponding chiral superfields while $\tilde{G}$ is the gluino
field.  The gluino mass is given by $m_{\tilde{G}}=|m_{G3}|$.  With
the mSUGRA boundary condition, the off-diagonal elements in the mass
matrix of $\tilde{\bar{D}}$ are approximately given by
\begin{eqnarray}
    [m^2_{\tilde{\bar{D}}}]_{ij} &\simeq&
    -\frac{1}{8\pi^2} \left[ Y_N^\dagger Y_N \right]_{ij}
    (3m_0^2+a_0^2) \log \frac{M_*}{M_{\rm GUT}}
    \nonumber \\ &\simeq&
    -\frac{1}{8\pi^2} e^{-i(\phi^{(L)}_i-\phi^{(L)}_j)}
    \sum_{k}
    y_{\nu_k}^2 \left[V_{L}^*\right]_{ki} \left[V_{L}\right]_{kj}
    (3m_0^2+a_0^2) \log \frac{M_*}{M_{\rm GUT}},
\label{m_dR}
\end{eqnarray}
and hence non-vanishing off-diagonal elements are generated due to the
neutrino Yukawa matrix. Furthermore, in general, these off diagonal
elements have unknown $O(1)$ phases.  For a more precise calculation,
we solve the RG equation numerically.  With $m_0\gg m_{G5}$ and $a_0$,
the ratios of the off-diagonal to the diagonal elements are
approximately given by
\begin{eqnarray}
    \left| \frac{[m^2_{\tilde{\bar{D}}}]_{13}}
        {[m^2_{\tilde{\bar{D}}}]_{11}} \right| \simeq 
    6\times 10^{-4} 
    \times \left( \frac{M_{\nu_R}}{10^{14}\ {\rm GeV}} \right),~~~
    \left| \frac{[m^2_{\tilde{\bar{D}}}]_{23}}
        {[m^2_{\tilde{\bar{D}}}]_{11}} \right| \simeq 
    2\times 10^{-2}
    \times \left( \frac{M_{\nu_R}}{10^{14}\ {\rm GeV}} \right),
\label{Nm_dR}
\end{eqnarray}
where we used the neutrino masses and mixings given in Eq.\ 
(\ref{VL_la}).  The neutrino Yukawa couplings are proportional to
$M_{\nu_R}^{1/2}$, and hence the RG induced off-diagonal elements are
approximately proportional to $M_{\nu_R}$.  Similarly, the top Yukawa
coupling generates those of the left-handed down-type squarks:
\begin{eqnarray}
    [m^2_{\tilde{Q}}]_{ij} \simeq
    -\frac{1}{8\pi^2} 
    y_t^2 \left[ V_{\rm CKM} \right]_{ti} 
    \left[ V^*_{\rm CKM} \right]_{tj} (3m_0^2+a_0^2)
    \left( 3 \log \frac{M_*}{M_{\rm GUT}}
        + \log \frac{M_{\rm GUT}}{M_{\rm weak}} \right),
    \label{m_dL}
\end{eqnarray}
where $M_{\rm weak}$ is the electroweak scale.  The phases in
$[m^2_{\tilde{Q}}]_{ij}$ is governed by that in the CKM matrix in the
mSUGRA case.  Numerically, we obtain
\begin{eqnarray}
    \left| \frac{[m^2_{\tilde{Q}}]_{13}}{[m^2_{\tilde{Q}}]_{11}} 
    \right| \simeq 
    2\times 10^{-3},~~~
    \left| \frac{[m^2_{\tilde{Q}}]_{23}}{[m^2_{\tilde{Q}}]_{11}} 
    \right| \simeq 
    9\times 10^{-3}.
\label{Nm_dL}
\end{eqnarray}
Since the off-diagonal elements $[m^2_{\tilde{\bar{D}}}]_{i3}$ and
$[m^2_{\tilde{Q}}]_{i3}$ are coefficients of $\Delta B\neq 0$
operators, they change the standard model predictions to the mixing
and decay of the $B$-mesons.  In Fig.\ \ref{fig:feyn}, we show the
Feynman diagrams contributing to $\Delta B=2$ and $\Delta B=1$
processes.

Before discussing the $B_d\rightarrow \phi K_S$ process, let us
consider $B_d\rightarrow \psi K_S$ which is the primary target of the
$B$-factories.  In the standard model, CP violation in this process is
from the $B_d$-$\bar{B}_d$ mixing (with the standard parametrization
of the CKM matrix).  Once the CP asymmetry in $B_d\rightarrow \psi
K_S$ is measured, the phase ${\rm arg}(-[V_{\rm CKM}]_{cd}[V_{\rm
CKM}^*]_{cb}/ [V_{\rm CKM}]_{td}[V_{\rm CKM}^*]_{tb})$ is determined.

If there is a new source of the flavor and CP violations, however,
this prediction is changed.  In the SUSY GUT with the right-handed
neutrinos, the sdown-sbottom mixing through
$[m^2_{\tilde{\bar{D}}}]_{13}$ and $[m^2_{\tilde{Q}}]_{13}$ becomes an
origin of the $B_d$-$\bar{B}_d$ mixing. The SUSY contribution to the
$B_d$-$\bar{B}_d$ mixing is discussed in Ref.\ \cite{NPB477-321} where
it was pointed out that $[m^2_{\tilde{Q},
\tilde{\bar{D}}}]_{13}/[m^2_{\tilde{Q}, \tilde{\bar{D}}}]_{11}$ has to
be larger than $O(10^{-2})$ to change the standard model prediction
significantly.  In the present case, these ratios are at most
$O(10^{-3})$ with the largest possible value of the right-handed
neutrino mass, $M_{\nu_R}\sim 10^{15}\ {\rm GeV}$.  Therefore, we
expect that the SUSY contribution to the $B_d$-$\bar{B}_d$ mixing is
smaller than the future experimental sensitivity.  Indeed, we
calculated $\Delta\phi^{\rm mix}_{{B}_d\rightarrow \bar{B}_d}$, the
SUSY contribution to the phase in the $B_d$-$\bar{B}_d$ mixing
amplitude.  We found that $\Delta\phi^{\rm mix}_{{B}_d\rightarrow
\bar{B}_d}$ is typically $O(0.1\ \%)$ of the standard model
contribution, which is too small to be seen in the experiments.

The CP violation in the process $B_d\rightarrow\psi K_S$ is also
affected if the phase in the decay amplitude is changed.  For the
process $B_d\rightarrow\psi K_S$, however, the standard model
contribution to the decay amplitude is at the tree level while the
SUSY ones are at the one-loop level.  As a result, the SUSY
contribution to the decay amplitude is negligible.  In this scenario,
the CP violation in the decay $B_d\rightarrow \psi K_S$ is likely to
be consistent with the standard model prediction.  For
$B_d\rightarrow\psi K_S$, the SUSY contribution is relatively
insignificant because this process has a tree level decay amplitude in
the standard model.

For processes without tree level decay amplitude, the SUSY
contribution may be more significant.  The process $B_d\rightarrow\phi
K_S$ is such a process \cite{PLB395-241,NPB508-3}.  At the quark
level, $\Delta B=1$ operators contributing to $B_d\rightarrow\phi K_S$
have a structure like $(\bar{s}b)(\bar{s}s)$.  Such operators are
induced only at the one-loop level in the standard model.

The SUSY contribution to the decay amplitude for $B_d\rightarrow\phi
K_S$ is from the penguin and box diagrams.  (See Fig.\ 
\ref{fig:feyn}.)  The $\Delta B=1$ effective Lagrangian has the
following form:
\begin{eqnarray}
    {\cal L}_{\rm eff} &=& 
    C_{RR} (\bar{s}^a \gamma^\mu P_R b^a)
    (\bar{s}^b \gamma^\mu P_R s^b)
    \nonumber \\ &&
    + C_{RL}^{V} (\bar{s}^a \gamma^\mu P_R b^a)
    (\bar{s}^b \gamma^\mu P_L s^b)
    + C_{RL}^{S} (\bar{s}^a P_R b^a) (\bar{s}^b P_L s^b)
    \nonumber \\ &&
    + m_b C_{R}^{\rm DM} T^A_{ab} 
    \bar{s}^a [\gamma^\mu, \gamma^\nu] P_L b^b G^A_{\mu\nu}
    + (L\leftrightarrow R) + {\rm h.c.},
\end{eqnarray}
where $m_b$ is the bottom quark mass, $G^A_{\mu\nu}$ is the gluon
field strength, $T^A_{ab}$ is the $SU(3)_C$ generator and the indices
$a$ and $b$ are the color indices.  We calculate the SUSY and the
standard model contributions to the coefficients.  In our analysis, we
only consider the dominant contribution from the squark-gluino loops,
and the expressions for the SUSY contribution are given in the
Appendix.  Then, we estimate the decay amplitude:
\begin{eqnarray}
    {\cal M}_{\bar{B}_d\rightarrow \phi K^0}
    = \langle \phi K^0 | {\cal L}_{\rm eff}
    | \bar{B}_d \rangle .
\end{eqnarray}
In our calculation, no QCD corrections below the electroweak scale are
included.  Then, we adopt the factorization approximation to obtain
\begin{eqnarray}
    \frac{{\cal M}_{\bar{B}_d\rightarrow \phi K^0}}
    {2(p_B \epsilon_\phi) m_\phi^2 f_\phi F_+^{BK}}
    &=& \frac{1}{4} \Bigg[ \left( 1+\frac{1}{N_C} \right) C_{RR} 
    + C_{RL}^{V} 
    - \frac{1}{2N_C} C_{RL}^{S} 
    \nonumber \\ &&
    + \frac{1}{2} \left( 1 - \frac{1}{N_C^2} \right)
    g_3 \kappa_{\rm DM} C_R^{\rm DM} \Bigg] 
    + (L\leftrightarrow R),
\end{eqnarray}
where $g_3$ is the $SU(3)_C$ gauge coupling constant, $N_C=3$, and the
following relations are used
\begin{eqnarray}
    \langle \phi (p_\phi, \epsilon_\phi) | \bar{s}^a \gamma^\mu s^a
    | 0 \rangle &=& 
    m_\phi f_\phi \epsilon_\phi^\mu,
    \\
    \langle K^0 (p_K) | \bar{s}^a \gamma^\mu b^a
    | \bar{B}_d (p_B) \rangle  &=& 
    F_+^{BK} (p_B + p_K)^\mu + F_-^{BK} (p_B - p_K)^\mu.
\end{eqnarray}
Furthermore, $\kappa_{\rm DM}$ is an $O(1)$ coefficient from the
hadronization of the chromo-dipole moment operator.  We estimate
$\kappa_{\rm DM}$ using relations derived from the quark model
\cite{PLB377-161} and the heavy quark effective theory
\cite{PRD42-2388}:
\begin{eqnarray}
    \kappa_{\rm DM} \simeq \frac{m_b^2}{2q^2} 
    \left[ \frac{9}{8} + O(m_\phi^2 / m_b^2) \right],
\end{eqnarray}
where $q$ is the momentum transfer in the gluon line.  Typically,
$q^2=\frac{1}{2}(m_B^2-\frac{1}{2}m_\phi^2+m_K^2)$ \cite{PLB377-161},
which gives $\kappa_{\rm DM}\simeq 1.2$ \cite{NPB508-3}. In our
following discussion, we will present results with several values of
$\kappa_{\rm DM}$ to show the uncertainty related to $\kappa_{\rm
DM}$.

Now, we discuss the SUSY contribution to the decay amplitude ${\cal
M}^{\rm (SUSY)}_{B_d\rightarrow\phi K^0}$.  There are two types of
contributions, i.e., one proportional to
$[m^2_{\tilde{\bar{D}}}]_{23}$ and the other proportional to
$[m^2_{\tilde{Q}}]_{32}$.  In the mSUGRA approach,
$[m^2_{\tilde{\bar{D}}}]_{23}$ is generated by the neutrino Yukawa
matrix and is proportional to $e^{i(\phi^{(L)}_3-\phi^{(L)}_2)}$. The
phase in the CKM matrix is independent of $\phi^{(L)}$'s in the SUSY
$SU(5)$ model.  Since the standard model contribution to the decay
amplitude ${\cal M}^{\rm (SM)}_{\bar{B}_d\rightarrow\phi K^0}$ is
proportional to $[V^*_{\rm CKM}]_{ts}[V_{\rm CKM}]_{tb}$, the
right-handed down-type squark contribution may have, in general, an
arbitrary phase relative to the standard model one.  On the contrary,
in mSUGRA, $[m^2_{\tilde{Q}}]_{32}$ is approximately proportional to
$y_t^2[V^*_{\rm CKM}]_{ts}[V_{\rm CKM}]_{tb}$.  Furthermore, with the
model parameters we will use below, $|[m^2_{\tilde{Q}}]_{32}|$ becomes
much smaller than $|[m^2_{\tilde{\bar{D}}}]_{23}|$ since $|[V_{\rm
CKM}]_{ts}|\ll |[V_L]_{32}|$.  Therefore, when $y_{\nu_3}\sim y_t$,
the SUSY contribution is dominated by the one proportional to
$[m^2_{\tilde{\bar{D}}}]_{23}$ and ${\cal M}^{\rm
(SUSY)}_{B_d\rightarrow\phi K^0}$ is approximately proportional to
$e^{i(\phi^{(L)}_3-\phi^{(L)}_2)}$.  Of course, if we consider
different model, $[m^2_{\tilde{Q}}]_{32}$ may also contribute.

The CP violation in the decay process $B_d\rightarrow \phi K_S$ is
determined by the sum of the mixing and decay phases:
\begin{eqnarray}
    \phi^{\rm total}_{B_d\rightarrow \phi K_S} &=& 
    \phi^{\rm mix}_{{B}_d\rightarrow \bar{B}_d}
    + \phi^{\rm decay}_{\bar{B}_d\rightarrow \phi K^0}
\nonumber \\
    &=& \phi^{\rm mix}_{{B}_d\rightarrow \bar{B}_d}
    + {\rm arg} 
    \left[ {\cal M}^{\rm (SM)}_{\bar{B}_d\rightarrow\phi K^0}
        + {\cal M}^{\rm (SUSY)}_{\bar{B}_d\rightarrow\phi K^0} 
    \right].
\end{eqnarray}
The phase in the mixing is universal for the two processes
$B_d\rightarrow \psi K_S$ and $B_d\rightarrow \phi K_S$.  In addition,
the standard model predicts very small decay phases for these decay
modes.  As a result, in the standard model, $\phi^{\rm
total}_{B_d\rightarrow \phi K_S}$ should be almost the same as the CP
violating phase measured in $B_d\rightarrow \psi K_S$.  However, in
the model we consider, the decay phase can be different in two cases.
In order to estimate the SUSY contribution to the decay phase, we
calculate the quantity
\begin{eqnarray}
    \Delta\phi^{\rm decay}_{\bar{B}_d\rightarrow \phi K^0}
    \equiv
    \tan^{-1} \left(
        \frac{ |{\cal M}^{\rm (SUSY)}_{\bar{B}_d\rightarrow\phi K^0}| }
        { |{\cal M}^{\rm (SM)}_{\bar{B}_d\rightarrow\phi K^0} | }
    \right).
\end{eqnarray}
Notice that $|{\cal M}^{\rm (SUSY)}_{\bar{B}_d\rightarrow\phi K^0}|$
is almost independent of the phases $\phi^{(L)}$'s, and that
$\Delta\phi^{\rm decay}_{\bar{B}_d\rightarrow \phi K^0}$ is the
maximal possible correction to the decay phase for a given set of
model parameters (except the GUT phases).  Such a maximal value is
obtained when the phases are chosen such that
$\phi^{(L)}_3-\phi^{(L)}_2\simeq {\rm arg}[{\cal M}^{\rm
(SM)}_{\bar{B}_d\rightarrow\phi K^0}]+\pi/2$.

\begin{figure}
\centerline{\epsfxsize=0.5\textwidth\epsfbox{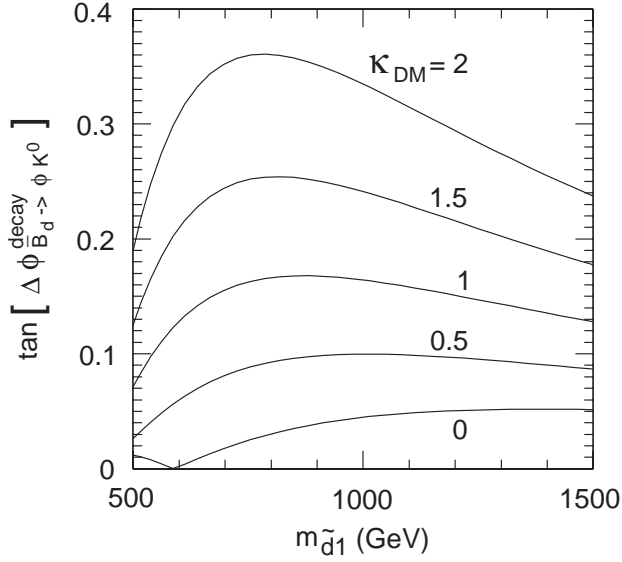}}
\caption{$\tan[\Delta\phi^{\rm decay}_{\bar{B}_d\rightarrow\phi
K^0}]$ as a function of the lightest down-type squarks mass with
$m_{\tilde{G}}=500\ {\rm GeV}$, $a_0=0$, $\tan\beta =3$,
$M_{\nu_R}=5\times 10^{14}\ {\rm GeV}$, and $\kappa_{\rm DM}$ is 0,
0.5, 1, 1.5, and 2 from below.}
\label{fig:dphi}
\end{figure}

In Fig.\ \ref{fig:dphi}, we plot $\tan[\Delta\phi^{\rm
decay}_{\bar{B}_d\rightarrow \phi K^0}]$ as a function of the lightest
down-type squark mass $m_{\tilde{d}1}$.  We plot the results with
$m_{\tilde{G}}={\rm 500\ GeV}$, $a_0=0$ and $\kappa_{\rm DM}=0$, 0.5,
1, 1.5, and 2.  Let us discuss the behavior of the SUSY contribution.
As mentioned in the Appendix, there are three classes of
contributions: box, color-charge form factor, and chromo-dipole
contributions.  We found that, within our approximation, box and
color-charge form factor contributions have almost the same size but
the opposite sign in most of the parameter space we studied.  As a
result, there is a significant cancellation between these two
contributions.  In particular, with the parameter we used for Fig.\ 
\ref{fig:dphi}, an exact cancellation occurs when
$m_{\tilde{d}1}\simeq 600\ {\rm GeV}$.  The chromo-dipole contribution
is comparable to the others when $\kappa_{\rm DM}\sim 1$.  However,
this contribution is very sensitive to $\kappa_{\rm DM}$, and hence
the final result strongly depends on the value of $\kappa_{\rm DM}$.
From Fig.\ \ref{fig:dphi}, we see that $\Delta\phi^{\rm
decay}_{\bar{B}_d\rightarrow \phi K^0}$ can be as large as $O(0.1)$
with the reasonable value of $\kappa_{\rm DM}\sim 1$, although more
precise calculation of $\Delta\phi^{\rm decay}_{\bar{B}_d\rightarrow
\phi K^0}$ requires better understanding of the hadronic matrix
elements.

Here, we should comment on the process $\mu\rightarrow e\gamma$.  With
the parameters we used, ${\rm Br}(\mu\rightarrow e\gamma)$ becomes
larger than the current upper bound $1.2\times 10^{-11}$ \cite{PDG}
when $m_{\tilde{d}1}\lesssim 900\ {\rm GeV}$.  However this is rather
a model-dependent statement since all the mixings in $V_L$ affect
${\rm Br}(\mu\rightarrow e\gamma)$.  If the 3-1 element of $V_L$ is
$\sim 0.1$, or if we adopt a non-universal right-handed neutrino mass
matrix, ${\rm Br}(\mu\rightarrow e\gamma)$ may become smaller.  In
addition, with the neutrino mass matrix suggested from the large angle
MSW solution with small mass splitting (i.e., so-called ``LOW''
solution) or with those suggested from the small angle MSW or vacuum
oscillation solutions to the solar neutrino problem (although they are
now statistically disfavored \cite{n2k_Suzuki}), ${\rm
Br}(\mu\rightarrow e\gamma)$ is more suppressed.  On the contrary,
$\Delta\phi^{\rm decay}_{\bar{B}_d\rightarrow \phi K^0}$ is sensitive
only to the 2-3 mixing in the neutrino sector.  Therefore, we do not
exclude the possibility of $m_{\tilde{d}1}\lesssim 900\ {\rm GeV}$ in
Fig.\ \ref{fig:dphi}.  Of course, even with $m_{\tilde{d}1}\gtrsim
900\ {\rm GeV}$, $\Delta\phi^{\rm decay}_{\bar{B}_d\rightarrow \phi
K^0}$ can be $O(0.1)$.

Since the standard model predicts almost the same mixing and decay
phases in the $B_d\rightarrow \psi K_S$ and $B_d\rightarrow \phi K_S$
processes, $\Delta\phi^{\rm decay}_{\bar{B}_d\rightarrow \phi K^0}\sim
O(0.1)$ should be an interesting signal.  Since the uncertainties are
expected to be $O(10^{-2})$ in the standard model calculation
\cite{PLB395-241}, $\Delta\phi^{\rm decay}_{\bar{B}_d\rightarrow \phi
K^0}\sim O(0.1)$ will be regarded as a sign of a new physics beyond
the standard model if observed.

So far, we have discussed the decay of the $B_d$-meson.  If the
$B_s$-meson is available, the similar analysis is possible.  The
discussion is almost parallel to the case of $B_d$-meson.  For the
process $b\rightarrow c\bar{c}s$ (resulting in, for e.g.,
$B_s\rightarrow\psi\eta, D_s\bar{D}_s$), a tree amplitude exists and
the SUSY contribution to the decay phase is negligible.  On the
contrary, $b\rightarrow s\bar{s}s$ (which induces
$B_s\rightarrow\phi\eta'$) occurs at the one-loop level and the SUSY
contribution can be significant.  In the standard model, very small CP
asymmetries are expected in these processes.  Once the SUSY
contribution is taken into account, however, the decay phase of
$O(0.1)$ may be induced for $B_s\rightarrow\phi\eta'$ since the
standard model contribution is one-loop suppressed.

In summary, the SUSY contribution may change the phase in
$B_d\rightarrow\phi K_S$. In the future, comparing this process with
$B_d\rightarrow\psi K_S$, we may see a difference in the decay phases
in these two processes, which cannot be explained by the standard
model.  This suggests the importance to study various decay modes of
the $B$-mesons.  Since the branching ratio of the process
$B_d\rightarrow\phi K_S$ is expected to be $O(10^{-5})$
\cite{PLB395-241}, such a study should be challenging at the first
stage of the present asymmetric $B$ factories.  However, at the second
stage, or at the hadron colliders, more $B$ samples are expected.
Then, the study of the decay modes with smaller branching ratios is an
interesting possibility to look for a signal from the new physics
beyond the standard model.  Therefore, it is desirable to collect a
large number of the $B$-mesons and to study the CP violation in the
various decay modes.

\section*{Acknowledgment}

The author would like to thank M.\ Tanabashi for helpful discussions.
This work is supported by the Grant-in-Aid for Scientific Research from
the Ministry of Education, Science, Sports, and Culture of Japan (No.\ 
12047201).

\appendix
\renewcommand{\theequation}{A.\arabic{equation}}
\setcounter{equation}{0}

\section*{Appendix: Coefficients}

In this appendix, we present the expressions for the SUSY contribution
to the coefficients of the $\Delta B=1$ operators:
\begin{eqnarray}
    {\cal L}_{\rm eff} &=& 
    C_{RR} (\bar{s}^a \gamma^\mu P_R b^a)
    (\bar{s}^b \gamma^\mu P_R s^b)
    \nonumber \\ &&
    + C_{RL}^{V} (\bar{s}^a \gamma^\mu P_R b^a)
    (\bar{s}^b \gamma^\mu P_L s^b)
    + C_{RL}^{S} (\bar{s}^a P_R b^a) (\bar{s}^b P_L s^b)
    \nonumber \\ &&
    + m_b C_{R}^{\rm DM} T^A_{ab} 
    \bar{s}^a [\gamma^\mu, \gamma^\nu] P_L b^b G^A_{\mu\nu}
    + (L\leftrightarrow R) + {\rm h.c.}
\end{eqnarray}
We only consider the dominant contribution from the squark-gluino
loops, and we use the mass-eigenstate basis.

With the soft SUSY breaking terms given in Eq.\ (\ref{L_soft}), the
mass matrix of the down-type squarks is given by\footnote{In Eq.\ 
(\ref{M_d^2}) expression, we omit the Yukawa and $D$-term
contributions to the diagonal terms, which are included in our
numerical calculation.}
\begin{eqnarray}
    {\cal M}^2_{\tilde{d}} = 
    \left( 
        \begin{array}{cc}
            {m^2_{\tilde{Q}}}^T &
            A_D^* \langle H_d\rangle + \mu Y_D^* \langle H_u\rangle
            \\
            A_D^T \langle H_d\rangle + \mu^* Y_D^T \langle H_u\rangle
            & {m^2_{\tilde{D}}}
        \end{array} \right),
    \label{M_d^2}
\end{eqnarray}
where $\mu$ is the SUSY invariant Higgs mass (i.e., so-called the
$\mu$-parameter).  The above mass matrix is diagonalized by the
unitary matrix $U_{\tilde{d}}$
\begin{eqnarray}
    [U_{\tilde{d}}^\dagger {\cal M}^2_{\tilde{d}} U_{\tilde{d}}]_{AB}
    = m^2_{\tilde{d}_A} \delta_{AB}.
\end{eqnarray}
Then, the coupling constant for the $d_i$-$\tilde{d}_A$-gluino vertex
is given by
\begin{eqnarray}
    X_{iA}^L = - \sqrt{2} g_3 [U_{\tilde{d}}]^*_{i,A} 
    e^{-i\phi_{\tilde{G}}},~~~
    X_{iA}^R = - \sqrt{2} g_3 [U_{\tilde{d}}]^*_{i+3,A} 
    e^{i\phi_{\tilde{G}}},
\end{eqnarray}
where $\phi_{\tilde{G}}$ is the phase in the gluino mass
\begin{eqnarray}
    m_{G3} = |m_{G3}| e^{-2i\phi_{\tilde{G}}} \equiv 
    m_{\tilde{G}} e^{-2i\phi_{\tilde{G}}}.
\end{eqnarray}
In our calculation, we take $\phi_{\tilde{G}}=0$ to evade the
constraint from the electric dipole moments.

Neglecting the left-right mixing in the down-type squark mass matrix,
the box contribution is
\begin{eqnarray}
    \left. \tilde{C}_{RR} \right|_{\rm Box} &=&
    \frac{1}{16\pi^2m_{\tilde{G}}^2} 
    \sum_{AB} X_{sA}^{R*}X_{bA}^{R} X_{sB}^{R*}X_{sB}^{R} 
    \nonumber \\ && \times
    \left[ \frac{N_C^3-2N_C+1}{4N_C^2} B_1(x_A,x_B)
        + \frac{N_C^2-2N_C+1}{8N_C^2} B_2(x_A,x_B)
    \right],
    \\
    \left. \tilde{C}_{RL}^V \right|_{\rm Box} &=&
    \frac{1}{16\pi^2m_{\tilde{G}}^2} 
    \sum_{AB} X_{sA}^{R*}X_{bA}^{R} X_{sB}^{L*}X_{sB}^{L} 
    \nonumber \\ && \times
    \left[ - \frac{N_C^2+1}{4N_C^2} B_1(x_A,x_B)
        - \frac{1}{8N_C^2} B_2(x_A,x_B)
    \right],
    \\
    \left. \tilde{C}_{RL}^S \right|_{\rm Box} &=&
    \frac{1}{16\pi^2m_{\tilde{G}}^2} 
    \sum_{AB}
    X_{sA}^{R*}X_{bA}^{R} X_{sB}^{L*}X_{sB}^{L} 
    \nonumber \\ && \times
    \left[ - \frac{1}{4N_C} B_1(x_A,x_B)
        + \frac{N_C^2-2}{4N_C} B_2(x_A,x_B)
    \right],
\end{eqnarray}
where the ``tilde'' denote the SUSY contribution to the coefficients.
Here, the functions $B_1$ and $B_2$ are given by
\begin{eqnarray}
    B_1(x_A,x_B) &=&  
%    16\pi^2m_{\tilde{G}}^2 \times \frac{1}{4} I_{14}(XXAB) 
%    \nonumber \\ &=&
    -\frac{x_A^2 \log x_A}{4(x_A-x_B)(x_A-1)^2}
    -\frac{x_B^2 \log x_B}{4(x_B-x_A)(x_B-1)^2}
    \nonumber \\ &&
    -\frac{1}{4(x_A-1)(x_B-1)},
    \\
    B_2(x_A,x_B) &=&
%    16\pi^2m_{\tilde{G}}^2 \times m_{\tilde{G}}^2 I_{04}(XXAB)
%    \nonumber \\ &=&
    -\frac{x_A \log x_A}{(x_A-x_B)(x_A-1)^2}
    -\frac{x_B \log x_B}{(x_B-x_A)(x_B-1)^2}
    \nonumber \\ &&
    -\frac{1}{(x_A-1)(x_B-1)},
\end{eqnarray}
with $x_A\equiv m_{\tilde{d}_A}^2/m_{\tilde{G}}^2$. Notice that,
approximately, the following relation holds in our case:
\begin{eqnarray}
    \sum_B X_{sB}^{L*}X_{sB}^{L} B_i(x_A,x_B)
    \simeq \sum_B X_{sB}^{R*}X_{sB}^{R} B_i(x_A,x_B)
    \simeq 2g_3^2 B_i(x_A,x_{\tilde{s}}),
\end{eqnarray}
where $B_i=B_1, B_2$.

The contribution from the color-charge form factor are
\begin{eqnarray}
    \left. \tilde{C}_{RR} \right|_{\rm CF} =
    \frac{N_C-1}{2N_C} \tilde{C}_R^{\rm CF},~~~
    \left. \tilde{C}_{RL}^V \right|_{\rm CF} =
    - \frac{1}{2N_C} \tilde{C}_R^{\rm CF},~~~
    \left. \tilde{C}_{RL}^S \right|_{\rm CF} =
    - \tilde{C}_R^{\rm CF},
\end{eqnarray}
where
\begin{eqnarray}
    \tilde{C}_R^{\rm CF} &=& 
    \frac{g_3^2}{16\pi^2 m_{\tilde{G}}^2} 
    \sum_{A}
    X_{sA}^{R*}X_{bA}^{R} \left[
    - \frac{1}{2N_C} C_1 (x_A) 
    + \frac{1}{2}N_C C_2 (x_A) \right],
\end{eqnarray}
with
\begin{eqnarray}
    C_1 (x) &=&
%    16\pi^2m_{\tilde{G}}^2 \times 
%    \left( \frac{-1}{6} I_{25}(XAAAA) \right)
%    \nonumber \\ &=& 
    \frac{2x^3-9x^2+18x-11-6\log x}{36(1-x)^4},
    \\
    C_2 (x) &=&
%    16\pi^2m_{\tilde{G}}^2 \times 
%    \left( \frac{-1}{3} I_{25}(XXXXA)
%    + \frac{1}{2} m_{\tilde{G}}^2 I_{15}(XXXXA) \right)
%    \nonumber \\ &=& 
    \frac{-16x^3+45x^2-36x+7+6x^2(2x-3)\log x}{36(1-x)^4}.
\end{eqnarray}

For the chromo-dipole operator,
\begin{eqnarray}
    \tilde{C}_R^{\rm DM}
    &=& \frac{g_3}{64\pi^2 m_{\tilde{G}}^2 m_b}
    \sum_A
    \Bigg[
    -\frac{1}{2N_C} \left\{ m_b X_{sA}^{R*}X_{bA}^{R} D_1 (x_A)
        + m_{\tilde{G}} X_{sA}^{R*}X_{bA}^{L} D_2 (x_A)
    \right\}
    \nonumber \\ &&
    + \frac{1}{2}N_C  \left\{ m_b X_{sA}^{R*}X_{bA}^{R}
        D_3 (x_A) 
        + m_{\tilde{G}} X_{sA}^{R*}X_{bA}^{L} D_4 (x_A)
    \right\}
    \Bigg]
\end{eqnarray}
where
\begin{eqnarray}
    D_1 (x) &=&
%    16\pi^2m_{\tilde{G}}^2 \times
%    \left( - m_A^2 I_{15}(AAAAX) \right) 
%    \nonumber \\ &=&
    \frac{-x^3+6x^2-3x-2-6x\log x}{6(1-x)^4},
    \\
    D_2 (x) &=&
%    16\pi^2m_{\tilde{G}}^2 \times \left( I_{14}(XXAA) \right)
%    \nonumber \\ &=&
    \frac{-x^2+1+2x\log x}{(x-1)^3},
    \\
    D_3 (x) &=&
%    16\pi^2m_{\tilde{G}}^2 \times 
%    \left( I_{25}(XXXXA) - I_{14}(XXXA) \right)
%    \nonumber \\ &=&
    \frac{2x^3+3x^2-6x+1-6x^2\log x}{6(1-x)^4},
    \\
    D_4 (x) &=&
%    16\pi^2m_{\tilde{G}}^2 \times 
%    \left( -2I_{14}(XXXA) \right)
%    \nonumber \\ &=&
    \frac{-3x^2+4x-1+2x^2\log x}{(x-1)^3}.
\end{eqnarray}

Other coefficients, $\tilde{C}_{LL}$, $\tilde{C}_{LR}^V$,
$\tilde{C}_{LR}^S$, and $\tilde{C}_L^{\rm DM}$, are obtained by
interchanging $L\leftrightarrow R$.

\end{document}